# Does the Web of Science Accurately Represent Chinese Scientific Performance?

**Fei Shu** [*]
Chinese Academy of Science and Education Evaluation (CASSE), Hangzhou Dianzi University
Xiasha, Hangzhou, Zhejiang, 310018, China P.R.
Email: fei.shu28@gmail.com

**Charles-Antoine Julien**
School of Information Studies, McGill University
3661 Peel St., Montreal, Quebec, H3A1X1, Canada
Email: charles.julien@mcgill.ca

**Vincent Larivière**
École de bibliothéconomie et des sciences de l'information, Université de Montréal
C.P. 6128, Succ. Centre-Ville, Montréal, QC. H3C3J7, Canada
Email: vincent.lariviere@umontreal.ca

---

[*] Corresponding Author

# Abstract

*With the significant development of China's economy and scientific activity, its scientific publication activity is experiencing a period of rapid growth. However, measuring China's research output remains a challenge since Chinese scholars may publish their research in either international or national journals, yet no bibliometric database covers both Chinese and English scientific literature. The purpose of this study is to compare Web of Science (WoS) with a Chinese bibliometric database in terms of authors and their performance, demonstrate the extent of the overlap between the two groups of Chinese most productive authors in both international and Chinese bibliometric databases, and determine how different disciplines may affect this overlap. The results of this study indicate that Chinese bibliometric databases, or a combination of WoS and Chinese bibliometric databases, should be used to evaluate Chinese research performance except in few disciplines in which Chinese research performance could be assessed using WoS only.*

## Introduction

Over the last 20 years, the contribution of China[1] to the world's scientific activity—as measured by its number of Web of Science (WoS) publications—has increased at an impressive rate (Zhou, 2013). While part of this trend might be attributed to an increase in the number of research papers written in English by Chinese researchers (Montgomery, 2013), some Chinese scholars might still prefer to publish their manuscripts in Chinese scholarly journals that are indexed by Chinese bibliometric databases only (Jin, Zhang, Chen, & Zhu, 2002; Moed, 2002b). Hence, measuring China's research output remains a challenge, as no bibliometric database covers both Chinese and English scientific literature.

Many scholars have concluded that the WoS is not an appropriate tool to measure Chinese research performance (Guan & He, 2005; Jin & Rousseau, 2004; Zhou & Leydesdorff, 2007), as significant differences have been found in the coverage of international and national Chinese bibliometric databases (Hennemann, Wang, & Liefner, 2011; Meho & Yang, 2007). While previous work has attempted to explain differences between WoS and Chinese bibliometric databases by looking at journal hierarchies and citation relations (Zhou & Leydesdorff, 2007), or regional publications (Liang, 2003), no research has yet analysed the discrepancies at the level of authors. For instance, little is known on the extent to which scholars from Chinese institutions publish their articles in international journals, or whether they give up publishing papers in Chinese in order to be more visible internationally. A better understanding of these trends might help to explain the differences between the international and national Chinese bibliometric databases.

The purpose of this study is to compare an international bibliometric database (i.e., WoS) with a national Chinese bibliometric database in terms of authors and their publications, demonstrate the extent of the overlap between the two groups of Chinese most productive authors in both international and Chinese bibliometric databases, and determine how different disciplines affect this overlap. The results of this study can reveal the extent to which international bibliometric databases can be used to evaluate Chinese national research production and performance as a whole, and in individual research disciplines.

## Literature Review
### Related Work

Since 1990s, China's international publication count has increased at an exponential rate of 20% annually (Kostoff, Briggs, Rushenberg, Eowles, et al., 2007). This makes China the fastest growing source of scholarly article publications and the second largest source country in terms of the number of articles published in WoS (National Science Board, 2018). A number of studies have attempted to characterize Chinese research achievements beyond the number of scholarly publications. Table 1 shows that several international and Chinese scholars have tracked China's publication records, measured China's collaborative networks, evaluated their academic journals and bibliometric databases, assessed Chinese university rankings, and analyzed their international visibility.

---

[1] In this study, China refers to the mainland China, which is the geopolitical area under the direct jurisdiction of the People's Republic of China excluding Hong Kong and Macau.

The results obtained by bibliometric studies depend on the chosen dataset. Table 1 presents the various databases used in these bibliometric studies; it shows that 59 out of 70 articles have used international bibliometric databases (i.e., WoS, Scopus) while 25 out of 70 articles used national Chinese bibliometric databases to evaluate Chinese research production, and that 14 studies (bold in Table 1) selected data from both international and national Chinese sources. This shows that international databases, especially WoS (used by 52 out of 70 articles), are still the major data sources for bibliometric studies on China. However, the extent to which they are representative of the Chinese output is not known. In other words, it is unclear whether the trends observed in international databases differ from those found in national Chinese databases. This study attempts to address this issue.

Table 1. Overview of Articles Using Bibliometric Databases in the Context of China

|  |  | General Evaluation of Chinese Research Performance | Collaboration Network | Journal & Database | University Ranking | International Visibility |
|---|---|---|---|---|---|---|
| International Bibliometric Database | WoS | **Ahlgren, Yue, Rousseau, and Yang (2017); Basu, Foland, Holdridge, and Shelton (2018)**; Gao and Guan (2009); Guan and He (2005); Jin and Rousseau (2004, 2005); Kostoff, Briggs, Rushenberg, Bowles, et al. (2007); Kostoff, Briggs, Rushenberg, Eowles, et al. (2007); Leydesdorff and Zhou (2005); Liang (2003); Liang, Havemann, Heinz, and Wagner-Döbler (2006); Liu, Tang, Gu, and Hu (2015); Mely, El Kader, Dudognon, and Okubo (1998); Meng, Hu, and Liu (2006); Moed (2002b); Zhi and Meng (2016); Zhou and Leydesdorff (2006); Zhou, Thijs, and Glänzel (2009a, 2009b) | **Hennemann et al. (2011)**; He (2009); Li and Li (2015); Niu and Qiu (2014); Park, Yoon, and Leydesdorff (2016); L. Wang, Wang, and Philipsen (2017); L. L. Wang and Wang (2017); X. Wang, Xu, Wang, Peng, and Wang (2013); Yuan et al. (2018); C. Zhang and Guo (2017); H. Zhang and Guo (1997); Z. Zhang, Rollins, and Lipitakis (2018); Zheng et al. (2012); Zhou and Glänzel (2010) | Basu (2010); Leydesdorff and Jin (2005); Liang (2003); Ren and Rousseau (2002); Shelton, Foland, and Gorelskyy (2009); S. Wang, Wang, and Weldon (2007); S. Wang and Weldon (2006); Zhou and Leydesdorff (2007) | Cheng and Liu (2008); Fu and Ho (2013); Liang, Wu, and Li (2001); Meho and Yang (2007); Qiu, Yang, and Zhao (2010) | **Basu (2010);** Fu, Chuang, Wang, and Ho (2011); Leydesdorff and Jin (2005); Qiu et al. (2010); Ren and Rousseau (2002); Shu and Larivière (2015); Shu, Lou, and Haustein (2018) |

|  | Scopus | **Basu, Foland, Holdridge, and Shelton (2018);** L. L. Wang (2016); | Royle, Coles, Williams, and Evans (2007); W. Wang, Wu, and Pan (2014) | **Basu (2010);** Ding, Zheng, and Wu (2012) | **Meho and Yang (2007);** Zhu, Hassan, Mirza, and Xie (2014) | **Basu (2010); Shu and Lariviere (2015);** L. L. Wang (2016) |
|---|---|---|---|---|---|---|
| National Chinese Bibliometric Database | **Chinese Science Citation Database (CSCD)** | **Ahlgren, Yue, Rousseau, and Yang (2017); Jin and Rousseau (2004); Liang (2003); Moed (2002)** | Liang and Zhu (2002) | **Leydesdorff and Jin (2005); Liang (2003);** Jin and Wang (1999); Rousseau, Jin, and Yang (2001) | **Liang, Wu, and Li (2001)** | **Leydesdorff and Jin (2005);** Rousseau et al. (2001) |
| | **Chinese Science and Technology Paper and Citation Database (CSTPCD)** | **Liang, Havemann, Heinz, and Wagner-Döbler (2006); Liang (2003); Guan and He (2005); Z. Wang, Li, Li, and Li (2012)** | Yan Wang, Wu, Pan, Ma, and Rousseau (2005) | Wu et al. (2004); **Zhou and Leydesdorff (2007);** | **Liang, Wu, and Li (2001)** | |
| | **China Academic Journals Full-Text Database (CJFD)** | Hu, Guo, and Hou (2017); **Z. Wang et al. (2012)** | | | | Yang, Ma, Song, and Qiu (2010) |
| | **Chinese Science and Technology Periodical Citation Database (VIP)** | **Z. Wang et al. (2012)** | **Hennemann et al. (2011)** | | | |
| | **Chinese Social Science Citation Index (CSSCI)** | Song, Ma, and Yang (2015) | Yan, Ding, and Zhu (2010) | | | |

Problem Statement

Bibliometric databases do not contain all of the published literature; they only represent the population of researchers they aim at studying (Okubo, 1997). Several studies have shown that the WoS may not adequately represent China's research activities (Guan & He, 2005; Jin & Rousseau, 2004; Jin et al., 2002; Liang et al., 2001; Moed, 2002b; Zhou & Leydesdorff, 2007), as more than 97% of Chinese language scholarly journals are excluded from its coverage (ISTIC, 2014). Moed (2002b) and Liang (2003) have suggested using national Chinese bibliometric databases to assess Chinese research performance. Some researchers have combined the WoS with national Chinese bibliometric databases and report many differences between the databases when investigating co-authorship networks (Hennemann et al., 2011), regional publications (Liang, 2003) as well as citation analysis (Meho & Yang, 2007). Taken as a whole, these results suggest that the WoS and national Chinese bibliometric databases tell different stories about Chinese research, although it is not clear how much they differ.

Critical factors to consider when analyzing data from different bibliometric databases are their coverage and comparability, which determine the study's validity and reliability (Hennemann et al., 2011). Previous studies show that coverage differences between WoS and national Chinese bibliometric databases will lead to different results (Liang, 2003; Zhou & Leydesdorff, 2007). It is not known to what extent all differences can be attributed to differences in coverage. This study addresses the lack of current research comparing coverage between WoS and a national bibliometric Chinese database at the level of individual authors.

In addition, since scholars in different disciplines have different traditions and habits of publication, publication activities vary significantly by discipline (Glänzel, 2003; Larivière, Archambault, Gingras, & Vignola-Gagnè, 2006; Okubo, 1997). Scholars in the Social Sciences and Humanities publish their works in books or monographs in addition to journals, through which scholars in the Natural Sciences diffuse most of their research findings. It is difficult to make comparisons of different disciplines due to the disciplinary variation (Okubo, 1997). Thus, this study compares two groups of Chinese most productive authors between WoS and a Chinese bibliometric database discipline by discipline.

### Research Questions

The purpose of this study is to compare the overlap in the Chinese most productive authors found in WoS and in a national Chinese bibliometric database, and describe the differences observed according to disciplines. It will answer the following research questions:

1. In a given discipline, to what extent are the most productive authors (in terms of numbers of publications) are same in both WoS and a Chinese database?
2. In a given discipline, to what extent are the institutional affiliations of these most productive authors are same in both WoS and a Chinese database?

## Methodology

### Research Design

#### Database

In this study, the Web of Science (WoS) and the Chinese Science and Technology Periodical Citation Database (VIP) are used as data sources because of their coverage and representation. WoS is the only bibliometric database covering a century of citation-based indicators for all disciplines, as well as, since 1973, all authors and their institutional affiliations (Moed, 2005). Indeed, most previous bibliometric studies on China are based on WoS (Zhou & Leydesdorff, 2007). Although WoS Categories are found to be problematic in terms of inappropriate journal classifications and multidisciplinary journal classification (Janssens, Zhang, De Moor, & Glänzel, 2009; L. Zhang, Janssens, Liang, & Glänzel, 2010), they are frequently used in research evaluation in China. There are five major bibliometric databases in China (see Table 2), VIP has the largest coverage and no obvious drawback; it also offers author rankings in terms of publications and citations that are not provided by other databases (Zhao, Lei, Ma, & Qiu, 2008). Thus, VIP was selected for the comparison with WoS in this study.

*Table 2. Comparisons of 5 Chinese Bibliometric Databases*

|  | Chinese Science Citation Database | Chinese Science and Technology Paper and Citation Database | Chinese Social Science Citation Index | China Academic Journals Full-Text Database | Chinese Science and Technology Periodical Citation Database |
|---|---|---|---|---|---|
| **Initial** | CSCD | CSTPCD | CSSCI | CJFD | VIP |
| **Chinese Name** | 中国科学引文数据库 | 中国科技论文与引文数据库 | 中文社会科学引文索引 | 中国学术期刊全文数据库 | 中文科技期刊引文数据库 |
| **URL** | http://sciencechina.cn | http://www.istic.ac.cn | http://cssci.nju.edu.cn/ | http://oversea.cnki.net/ | http://www.cqvip.com/ |
| **Coverage (in 2017)** | 1,195 journals | 2,054 journals | 533 journals | 10,324 journals | 14,352 journals |
| **Established** | 1989 | 1987 | 1998 | 1994 | 1994 |
| **Update Frequency** | Yearly | Yearly | Yearly | Monthly | Quarterly |
| **Why not select?** | No coverage in Social Sciences and Humanities | Problematic journal selection criteria, database architecture, and keyword choices (Wu et al., 2004) | No coverage in Natural Sciences | Very limited citation data | N/A |

The VIP was established by the CQVIP Corporation in 1994. VIP indexes about 14,000 academic journals covering all disciplines, more than any other Chinese bibliometric database. VIP offers bibliometric indicators that measure Chinese scientific research performance in terms of the number of publications and citations by authors, institutions, journals or topics. which are not provided in other databases.

### Disciplinary Classification

WoS and VIP use different disciplinary classification systems. WoS assigns journals to 232 subject categories (disciplines) while the VIP classifies Chinese literature into 35 fields (major disciplines) and 457 subfields (disciplines) using Chinese Library Classification Scheme (Zhongguo Tushuguan Fenleifa [Chinese Library Classification], 2010). Even so, Chinese Library Classification Schemes could be converted to the corresponding WoS categories via a conversion table, which is used by Clarivate Analytics connecting WoS Chinese products (e.g. CSCD) to WoS core collection (e.g. SCI/SSCI/AHCI).

Equivalences between the WoS and VIP disciplinary classification systems were first established based on the descriptions of each subject category. Next, the results were confirmed by consulting the experts from Chinese Academy of Sciences (CAS) and Clarivate Analytics. Finally, this produced 116 obvious one-to-one matches. *Dance* was removed from the list since no Chinese publication was found in this WoS category. Therefore, 115 disciplines with equivalent classes across WoS and VIP were compared in this study (See Appendix I), which account for 66.08 % of Chinese publications (959,728 of 1,452,380) in WoS and 65.15% of literature (19,472,497 of 29,889,566) in VIP. This list includes 83, 21 and 12 disciplines[2] in Natural Sciences, Social Sciences, and Arts and Humanities respectively.

## Data

### Data Collection

All papers with a Chinese address (CU = Peoples R China) published between 2008 and 2015 (n=1,452,380) as well as their bibliographic information were retrieved from WoS and assigned to relevant disciplines. In the 115 selected disciplines, Chinese authors contributed the most papers in *Chemistry, Physics* (92,342), followed by *Engineering, Electrical & Electronic* (70,318) and *Optics* (49,038) while they only contributed 2, 5 and 6 papers in *Folklore*, *Literary Theory & Criticism* and *Film, Radio & Television* respectively. On the other hand, 29,940,090 Chinese papers published between 2008 and 2015 were indexed by VIP under 457 subfields (disciplines), ranging from 1,667 papers in *Physics, Condensed Matter* to 4,223,457 papers in *Education & Educational Research* In the 115 selected disciplines. No correlation (r=0.0131) was found between WoS and VIP in terms of the number of publication among these 115 disciplines.

In each discipline, Chinese authors were ranked by their number of published papers during the period of 2008-2015 in both WoS and VIP dataset. The top 100 (and tied) authors in the 115 disciplines were retrieved and formed 115 pairs of author groups, for a total of 26,969 records in the two databases.

### Author Name Disambiguation

Author name ambiguity is a significant issue when conducting bibliometric analysis at the level of individual researchers (Moed, 2002a). This is even more evident in studies that investigate Chinese and Korean names (Strotmann & Zhao, 2012). Although WoS indexes the complete first name of the authors from 2008 onwards, author name ambiguity remains an issue in WoS, especially since different Chinese names can be transliterated to a single English name. The issue of author name ambiguity is less important in the VIP data, as full author names are recorded using Chinese characters. However, there remain cases where Chinese authors share the same Chinese name.

Both automatic and manual validation were performed to disambiguate author names in the WoS and the VIP data. A combination of the author's full name and her/his primary institutional affiliation was used for automatic validation. A pilot test with fully manual validation was conducted based on data from 10 selected disciplines (Shu, Larivière, & Julien, 2016), and the results indicated that the automatic validation allows to disambiguate about 97% of WoS data and almost all VIP data. Exceptional cases were caused by two or more Chinese authors that share the same (Chinese or English) name, and who were active within the same institution or the same discipline. In addition to the automatic validation, a thorough manual validation (that lasted about 6 months) was performed to disambiguate these exceptions. In each discipline, the same name affiliated to different institutions was validated as either an author having multiple affiliations or different authors sharing the same name. Incomplete entries and inconsistent formats were also corrected. The manual validation disambiguated 120,953 ambiguous records regarding Chinese author names.

---

[2] History is classified as discipline under both Social Science and Arts and Humanities.

Meanwhile, in addition to typos and incomplete entries, serious institutional name ambiguity was also found in WoS data. For example, *JINAN-UNIV* refers to Jinan University located at city of Guangzhou in the province of Guangdong while *UNIV-JINAN* refers University of Jinan located in the city of Jinan in the province of Shandong; *BEIJING-UNIV-TECHNOL* (Beijing University of Technology) and *BEIJING-INST-TECHNOLOGY* (Beijing Institute of Technology) are two different institutions while both *BEIJING-INST-CHEM-TECHNOL* and *BEIJING-UNIV-CHEM-TECHNOL* refer to the same Beijing University of Chemical Technology (formerly Beijing Institute of Chemical Technology). Both *CHINESE-ACAD-MED-SCI* and *PEKING-UNION-MED-COLL* refer to the same institution with two different names (Chinese Academy of Medical Science and Beijing Union Medical College). Institution name disambiguation was conducted manually at the same time as the author name disambiguation was performed, and clarified 1,398 ambiguous records regarding Chinese institution names.

### Classification into institutional sectors

All institutional affiliations were classified into five sectors: universities, scientific institutes, enterprises, hospitals not affiliated with universities, and other sectors. Since universities play the dominant role in China's scientific research output, contributing 82.8% of monographs and 73.4% of journal articles including 83.0% of WoS papers (National Bureau of Statistics of China, 2015), all Chinese universities were further classified into two sub-categories: tier-1 universities and tier-2 universities that are defined by Ministry of Education of China (2016)[3].

### Indicators

In this study, Chinese authors were defined as those whose primary affiliated institution is in China, regardless of their citizenship. In WoS, all articles with a Chinese address were selected.

---

[3] There are 2,631 higher education institutions in China, including 1,243 universities offering undergraduate programs. Traditionally these universities can be classified into elite universities (Tier-1) and non-elite universities (Tier-2) by Ministry of Education of China. Most Tier-1 universities are owned by Ministry of Education or other ministries of the central government while most Tier-2 universities are owned by local government. Tier-1 universities have priorities over Tier-2 universities to admit talent students; they also could secure more financial and benefit from preferential policies comparing to non-elite universities.

The authors of these selected articles were likely to qualify as Chinese authors, but co-author(s) whose affiliated institution was not located in China were excluded. Authors with multiple affiliated institutions were manually validated for their eligibility.

For each of the 115 disciplines chosen, the number of papers per author was compiled in order to produce ranked lists of top Chinese authors in WoS and VIP. The top 100 (and tied) authors in terms of the number of publications produced between 2008 and 2015 in the 115 identified disciplines formed 115 pairs of Chinese most productive authors. The amount of overlap between each of these 115 sets of researchers indicated whether the Chinese most productive authors found in the WoS is the same as the one found in the VIP. For each discipline, the overlap between those researchers who are among the top 100 in WoS and the top 100 in VIP (hereafter referred to as the overlap rate) was calculated based on the formula,

$$\text{Overlap rate} = \frac{N}{\min(100, T_v, T_w)}$$

where N=number of shared Chinese most productive authors in both databases, $T_V$=number of top 100 and tied VIP authors, and $T_W$=number of top 100 and tied WoS authors. The denominator here should be the lowest value among 100, Tv and Tw.

For example, the overlap rate is 20% (21/105) when 21 shared authors are found between 105 authors in WoS and 115 authors in VIP; in another example, the overlap rate is 10% (10/100) as 10 shared authors are found between 50 authors in WoS and 120 authors in VIP.

The publication counts presented in this paper were based on the number of research articles and review articles but exclude editorials, book reviews, letters to the editor and meeting abstracts that are not generally considered original contributions to scholarly knowledge (Moed, 1996). In China, not all co-authorship credits are assigned based on an individual's scientific contribution but on the basis of seniority (Shen, 2016). However, Chinese bibliometric databases, including VIP, give full credit to all co-authors when counting the number of publications. This study applied the same approach regardless of the argument on whether a full count or divided count is better to measure the co-authorship.

In addition to the overlap rate, eight indicators were also compiled for each discipline for the purpose of data analysis, as shown in Table 3.

*Table 3. List of Indicators Used in Data Analysis*

| Indicator | Description |
|---|---|
| The Overlap Rate | The share of Chinese most productive authors found in both databases |
| The number of VIP papers | The number of papers that were published between 2008 and 2015 and indexed by VIP |
| The number of VIP authors | The number of Chinese scholars who published at least one paper indexed by VIP between 2008 and 2015 |
| The number of Chinese WoS papers | The number of papers that were published by Chinese scholars between 2008 and 2015 and indexed by WoS |
| The number of WoS papers | The number of papers that were published between 2008 and 2015 and indexed by WoS |
| The number of Chinese WoS authors | The number of Chinese scholars who published at least one paper indexed by WoS between 2008 and 2015 |
| The ratio of Chinese WoS papers to all WoS papers (Ratio$_{c2w}$) | The share of Chinese WoS papers to all WoS papers |
| The ratio of Chinese WoS papers to all Chinese papers (Ratio$_{cw2c}$) | The share of Chinese WoS papers to all Chinese papers including both WoS papers and VIP papers |
| The ratio of Chinese WoS authors to VIP authors (Ratio$_{w2v}$) | The ratio of the number of Chinese WoS authors to the number of VIP authors |

# Results

Among the 26,969 records retrieved from WoS and VIP (14,911 records from WoS and 12,058 records from VIP), 12,270 and 11,066 Chinese most productive authors as well as their primary affiliated institutions were identified from WoS and VIP, respectively, across the 115 selected disciplines. As noted above, Chinese most productive authors in multiple disciplines tied for the top 100 ranking. In addition, the total numbers of Chinese most productive authors in 7 disciplines in WoS and 3 disciplines in VIP totaled fewer than 100 because fewer than 100 Chinese authors published papers in these disciplines between 2008 and 2015.

As Figure 1 and Figure 2 show, the average overlap rate between the two groups of Chinese most productive authors was 11.36% ranging from 0% to 34.65% across the 115 disciplines. The overlaps in the Natural Sciences including Life Sciences & Biomedicine, Physical Sciences, and Technology[4] were higher than those in the Social Sciences and Humanities. Although the size of discipline in terms of the total number of publications varies, no correlation (r=0.0141) was found between the size of discipline and the overlap rate.

---

[4] WoS classifies research areas into five domains including Arts and Humanities, Social Science, Life Science & Biomedicine, Physical Science, and Technology, the last three of which constitute Natural Sciences. All WoS categories could correspond to research areas and be assigned to these five domains using a conversion table.

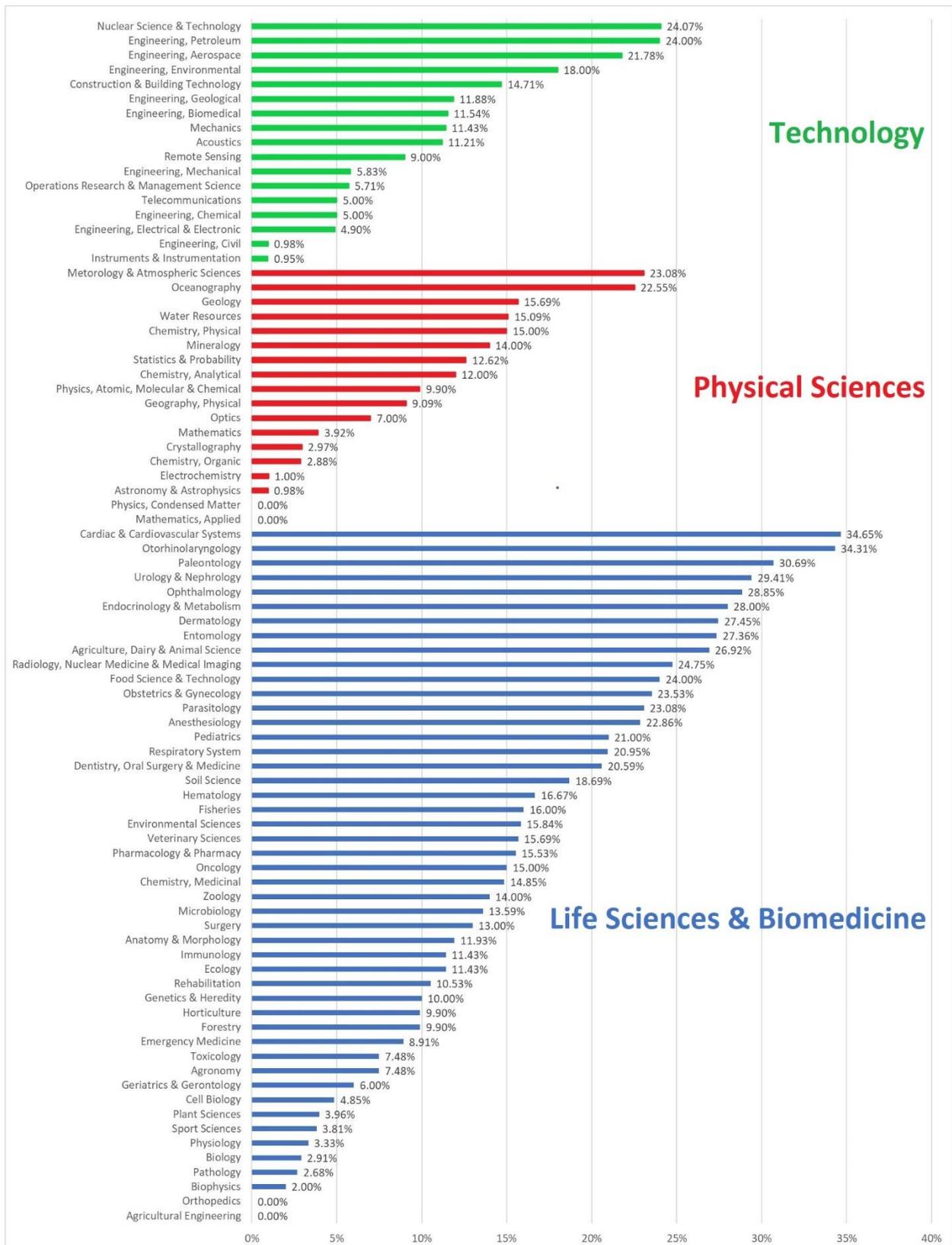

Figure 1. The Overlap Rate between Groups of Chinese Most Productive Authors in Natural Sciences

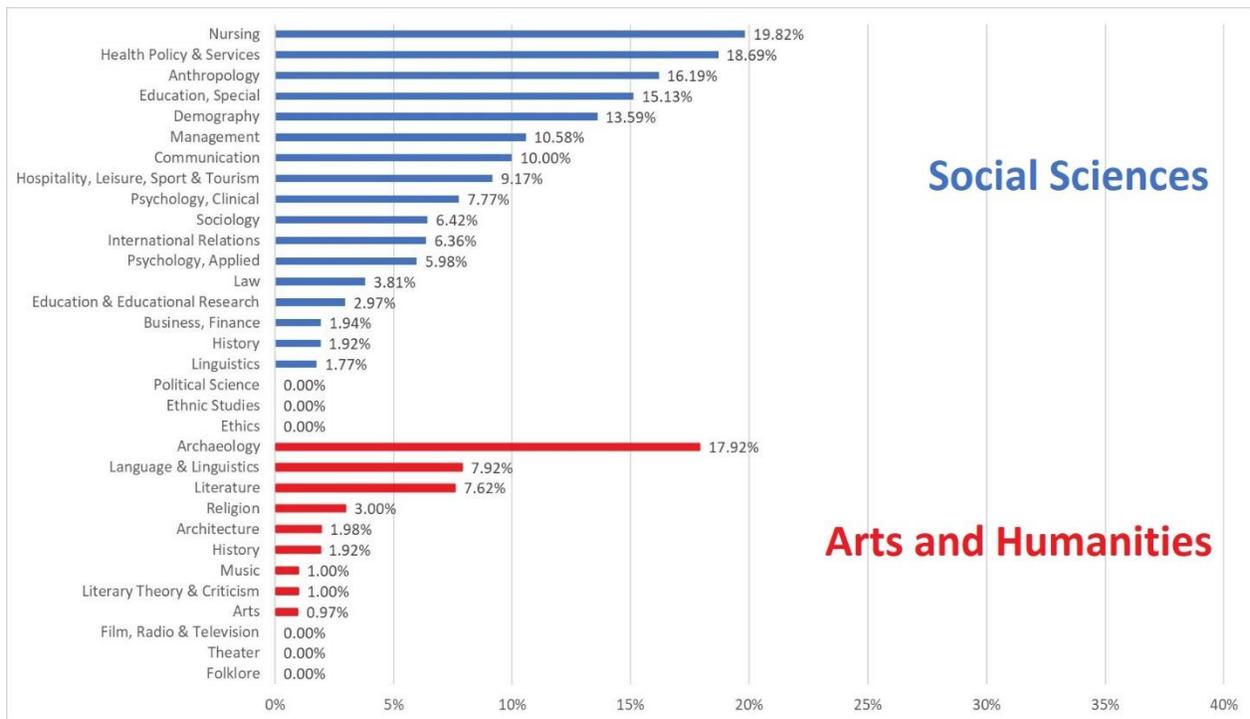

*Figure 2. The Overlap Rate between Groups of Chinese Most Productive Authors in Social Sciences and Humanities*

### Social Sciences and Humanities

Among the 13 disciplines in Arts and Humanities selected for this study, the average overlap rate was only 3.61%, which was mostly contributed by *Archaeology* (17.92%), where 19 authors were presented in both most productive author groups in WoS (148) and VIP (106). Indeed, during the period of 2008-2015, Chinese scholars only published 3,929 WoS papers in these 13 disciplines, ranging from 2 papers in *Folklore* to 1,203 papers in *Literature.* However, while Chinese scholars' contribution to WoS literature in Arts and Humanities remains marginal, they published 4,330,239 Chinese papers indexed by VIP in these 13 disciplines. The ratio of Chinese WoS papers to all Chinese papers (Ratio$_{cw2c}$) in these 13 disciplines was 0.09%. Few Chinese scholars published WoS papers in these disciplines, ranging from 2 authors in *Folklore* to 796 authors in *Archaeology*. Additionally, less than 100 Chinese authors published WoS papers during the period of 2008-2015 in 6 out of these 13 disciplines.

The overlaps were a little higher (7.61% in average) among the 21 disciplines in the Social Sciences, and ranged from 0% in *Ethics*, *Ethnic Studies*, and *Political Science* to 19.82% in *Nursing* which is related to Medical Sciences but classified as a Social Science discipline in WoS. Indeed, the top 3 Social Science disciplines in terms of the overlap (*Nursing*, *Health Policy & Services* and *Anthropology*) were all related to Health.

Compared to Chinese scholars in Arts and Humanities, Chinese scholars in Social Sciences published more WoS papers (20,507 across 21 disciplines), contributing 2.91% of international scientific production. However, the ratio of Chinese WoS papers to all Chinese papers (Ratio$_{cw2c}$) in Social Sciences remains very low (0.23%); in other words, more than 99% of Chinese papers in Social Sciences are published in National Chinese journals. The only exception was *Psychology, Applied*, where Chinese scholars published 948 WoS papers and 7,048 VIP papers respectively, but it is a small discipline considering that only 24,938 papers were indexed by WoS over the eight years. In addition, the number of Chinese authors in WoS's Social Sciences was also low, ranging from 21 in *Ethnic Studies* to 3,688 in *Management*.

### Natural Sciences

The overlap rates (13.24% on average) were higher in Natural Sciences than those in Social Sciences and Humanities. The overlap rates varied across the 83 disciplines, which could be classified into three broad categories in WoS: Life Sciences & Biomedicine, Physical Sciences, and Technology. In Life Sciences & Biomedicine, the average overlap rate was 15.54% ranging from 0% in *Orthopedics* and *Agricultural Engineering* to 34.65% in *Cardiac & Cardiovascular Systems, which* was the highest among all 115 disciplines. The average overlap rate was 9.32% in Physical Sciences ranging from 0% in *Physics, Condensed Matter* and *Mathematics, Applied* to 23.08% in *Meteorology & Atmospheric Sciences;* and the average overlap rate was 10.94% in Technology ranging from 0.95% in *Instruments & Instrumentation* to 24.07% in *Nuclear Science & Technology.*

The share of Chinese WoS papers to all WoS papers ($Ratio_{c2w}$) was higher in Natural Sciences than Social Sciences and Humanities. Chinese scholars contributed 14.25% of WoS papers during the period of 2008-2015 among these 83 disciplines in Natural Sciences (9.82% in Life Sciences & Biomedicine, 20.32% in Physical Sciences, and 18.64% in Technology) ranging from 2.22% in *Sports Science* to 31.00% in *Crystallography.* The correlation between the overlap rate and China's share in international scientific publications ($Ratio_{c2w}$) seemed to be negative (r=-0.4301) in Natural Sciences as shown in Figure 3, although a low correlation was observed in the interval of 10% to 20% on the X axis. The overlap rates were not higher among those disciplines in which Chinese scholars contributed more in international scientific literature.

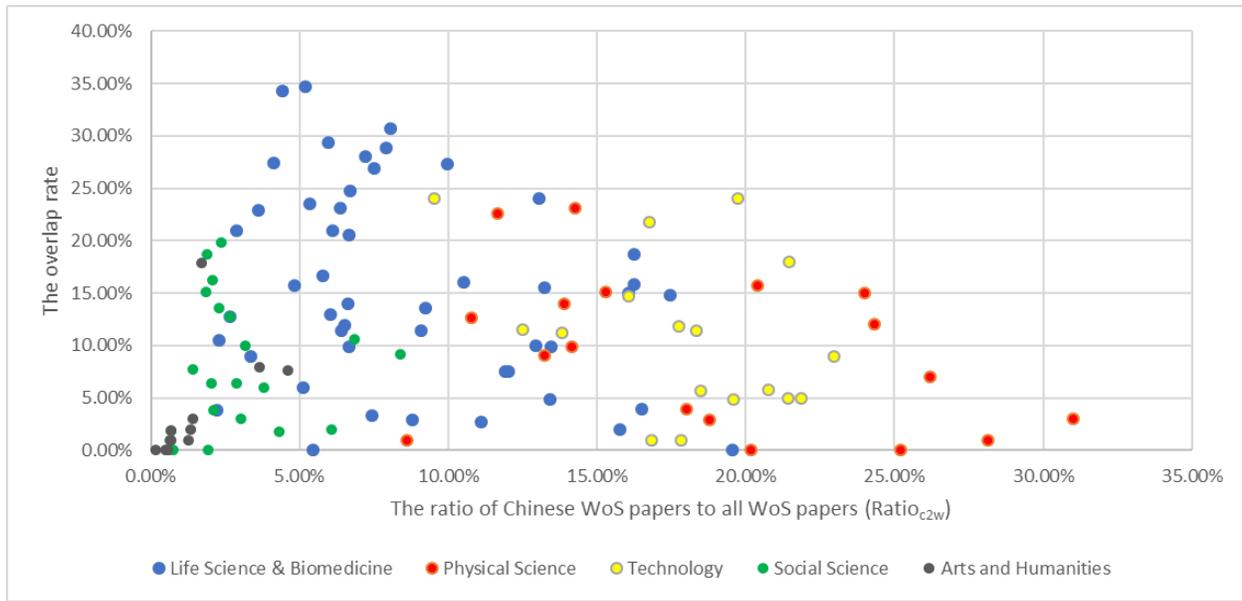

Figure 3. Scatterplot of the Overlap Rate and China's Share in International Scientific Literature

A negative correlation (r=-0.4409) was also found between the overlap rate and the ratio of Chinese WoS papers to all Chinese papers ($Ratio_{cw2c}$) in Natural Sciences. As shown in Figure 4, the overlap rates were low among those disciplines in which Chinese scholars published more international papers indexed by WoS. Indeed, the average $Ratio_{cw2c}$ was 12.16% among 83 disciplines in Natural Sciences (7.74% in Life Sciences & Biomedicine, 35.43% in Physical Sciences, and 7.35% in Technology respectively) ranging from 0.35% in *Sports Science* to 96.29% in *Physics, Condensed Matter.* In addition, disciplines in Social Sciences and Arts and Humanities have less variation than disciplines in Natural Sciences as shown in both Figure 3 and Figure 4.

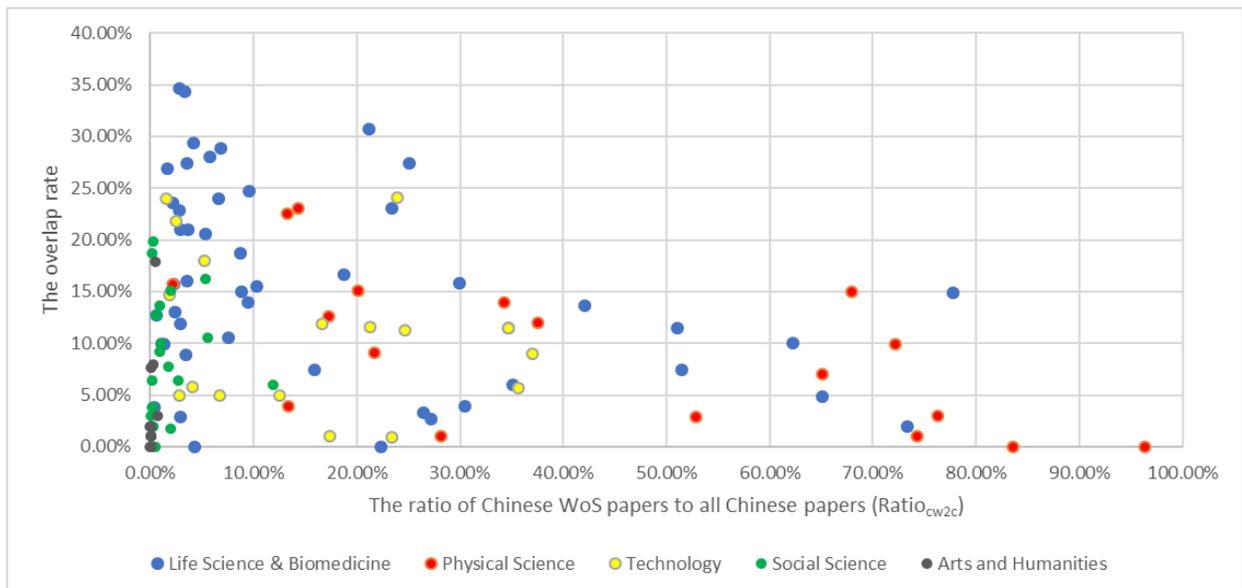

Figure 4. Scatterplot of the Overlap Rate and the Proportion of WoS Papers to China's National Scientific Literature

An interesting pattern was revealed when we investigated the relationship between the overlap rate and the combination of $Ratio_{c2w}$ and $Ratio_{cw2c}$. As Figure 5 shows, the overlap rates were less than 15% in all disciplines in which the $Ratio_{cw2c}$ was over 30%. When the threshold was increased to $Ratio_{cw2c} > 40\%$ and $Ratio_{c2w} > 10\%$, the overlap rates in 11 out of 13 disciplines within this section were less than 10%. It means that there are two different groups of most productive authors between WoS and VIP among those disciplines in which Chinese scholars contributed more in international scientific literature.

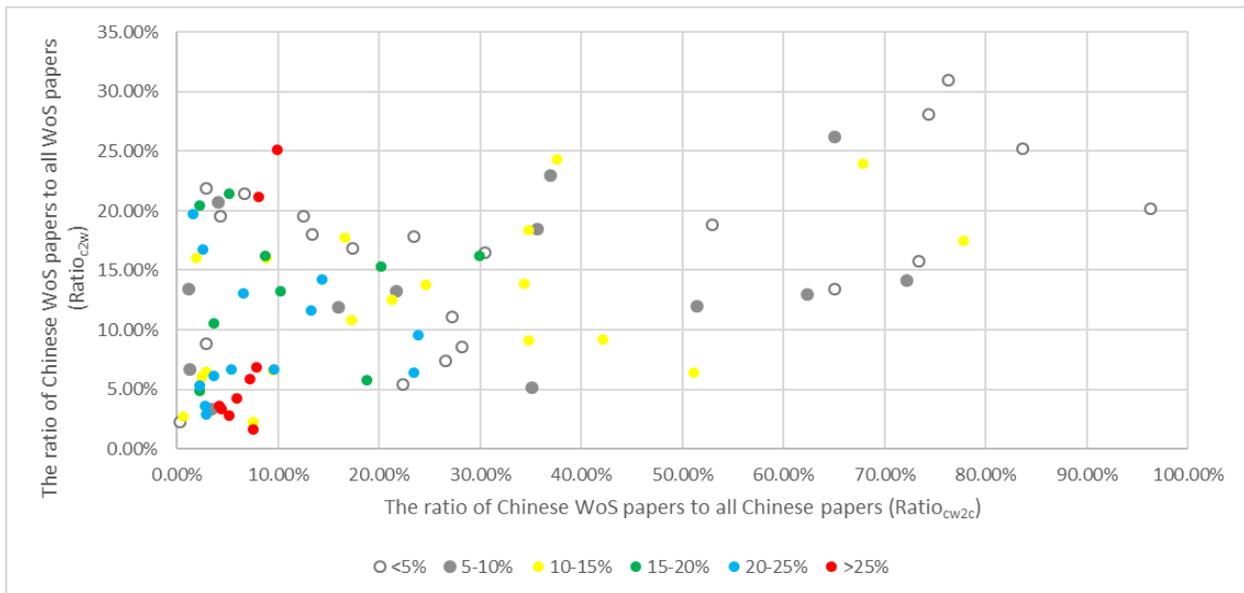

Figure 5. Scatterplot between the Overlap Rate and the Proportion of China's WoS papers to International and National Scientific Literature

### Affiliated Institution

As shown in Figure 6, the 11,066 Chinese most productive authors identified in VIP are from different sectors: universities contributed the most these authors (73.78%) including 40.44% from tier-1 universities and 33.35% from tier-2 universities, followed by scientific institutes (12.05%), non-affiliated hospitals (7.15%), Other sections (3.24%), Government (3.15%) and Enterprises (0.62%).

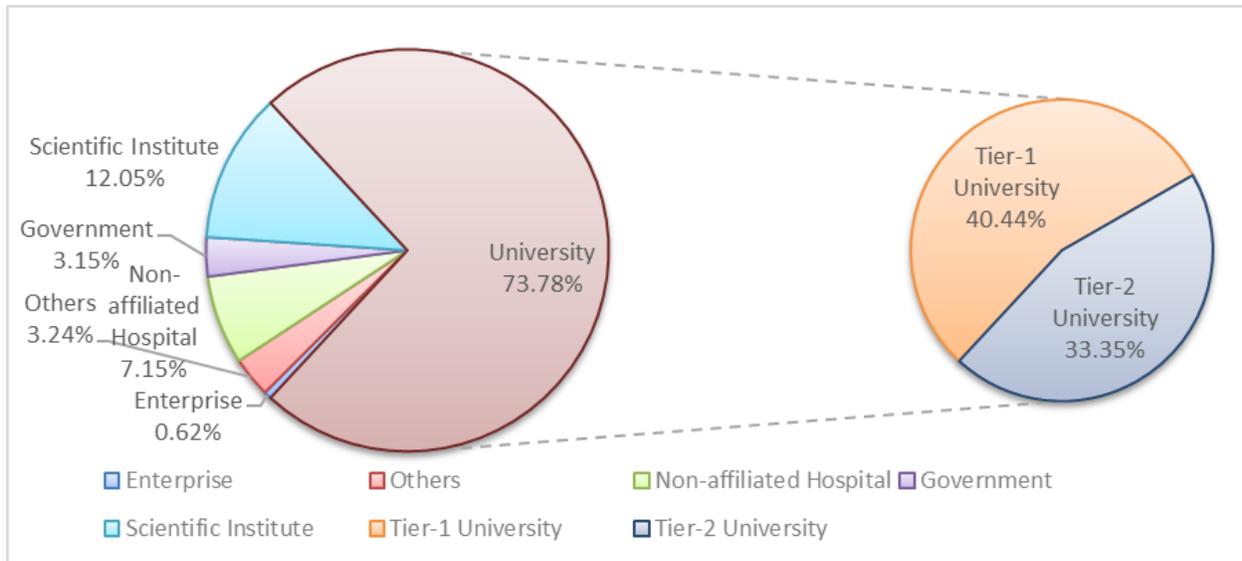

*Figure 6. Distribution of Chinese Most Productive Authors (VIP) by Type of Affiliated Institutions*

On the other hand, the distribution of Chinese most productive authors identified in WoS was slightly different; as shown in Figure 7, scientific institute contributed a similar share of most productive authors as 12.92%; the university contributed 81.64% of most productive authors while the contribution of non-affiliated hospital (3.06%), other sections (0.99%), government (0.80%) and enterprise (0.59%) were less than 5%. Indeed, 65.22% of most productive authors come from tier-1 universities while 16.42% of them were from tier-2 universities, which was significantly different from the ratio found in VIP.

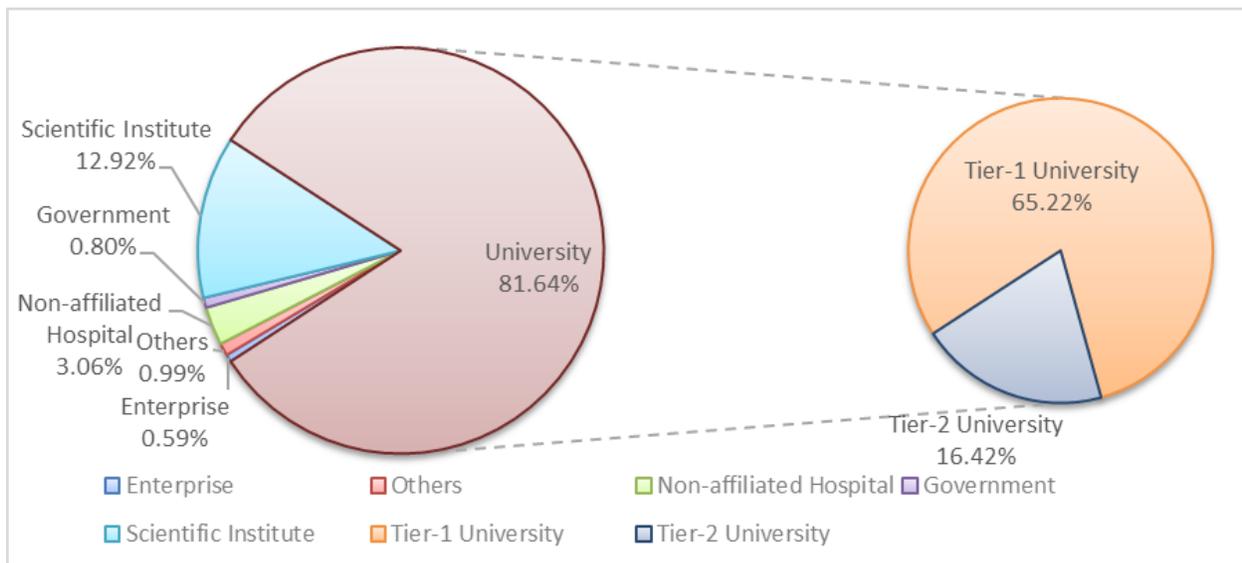

*Figure 7. Distribution of Chinese Most Productive Authors (WoS) by Type of Affiliated Institutions*

# Discussion

## Exceptional Disciplines

This study indicated that disciplines in Natural Sciences including Medical Science and Engineering exhibited a much higher level of overlap than those in Social Sciences and Humanities. It is unsurprising that scholars could easily disseminate knowledge internationally in the Natural Sciences in which all scholars share the same paradigm (Kuhn, 2012). On the other hand, scholars in Social Sciences and Humanities have to apply multi-paradigmatic approaches to understand complex social or human behaviour, making it a more difficult task to publish research in different languages in these disciplines (Cole, 1975; Delanty, 2005).

Although Chinese scholars contributed more international publications in Natural Sciences, we found an unexpected negative correlation between the overlap rate and both the ratio of Chinese WoS papers to all WoS papers ($Ratio_{c2w}$) and the ratio of Chinese WoS papers to all Chinese papers ($Ratio_{cw2c}$) in those disciplines. Low overlap rates of Chinese most productive authors between WoS and VIP were found in disciplines that are most international in scope. As shown in Figure 5 above, the overlap rates were less than or equal to 10% in 11 out of 13 disciplines in which the ratio of Chinese WoS papers to all WoS papers ($Ratio_{c2w}$) was over 10% and the ratio of Chinese WoS papers to all Chinese papers ($Ratio_{cw2c}$) was over 40%. The overlap rate declined to less than 3% in four disciplines in which the two ratios were increased to 20% and 70%, respectively, as shown in Table 4. Indeed, in those disciplines that were most international in scope, most Chinese scholars preferred diffusing their research results in international journals to publishing in national Chinese journals. For example, in *Physics, Condensed Matter*, 97,483 Chinese scholars published 43,319 WoS articles while 3,568 Chinese scholars published 1,667 articles in Chinese journals during the same period; Chinese scholars had almost abandoned publishing in national Chinese journals as 96.29% of their publications were in WoS journals. Thus, although the overlap rates were low, Chinese WoS papers could still represent Chinese research performance in those disciplines in which international publication was dominant.

*Table 4. Top 4 Disciplines that are Most International in Scope*

| Discipline | # VIP Papers | # VIP Authors | # WoS Papers | # WoS Authors | $Ratio_{c2w}$ | $Ratio_{cw2c}$ | $Ratio_{w2v}$ | Overlap Rate |
|---|---|---|---|---|---|---|---|---|
| **Physics, Condensed Matter** | 1,667 | 3,568 | 43,319 | 97,483 | 20.18% | 96.29% | 27.32 | 0.00% |
| **Mathematics, Applied** | 9,311 | 14,423 | 47,499 | 23,896 | 25.20% | 83.61% | 1.66 | 0.00% |
| **Crystallography** | 6,570 | 13,772 | 21,102 | 40,742 | 31.00% | 76.26% | 2.96 | 2.97% |
| **Electrochemistry** | 9,206 | 10,792 | 26,621 | 60,194 | 28.14% | 74.30% | 5.58 | 1.00% |

## Publication Patterns

Being a most productive scholar in any disciplines is competitive considering the pareto distribution of author productivity (Lotka, 1926). Being a Chinese most productive author in both Chinese publishing (VIP) and international publishing (WoS) means that the Chinese scholar has to allocate her/his manuscripts to two directions; some are sent to national journals while others are submitted to international journals. She/he also needs to balance the number of submission between national publication and international publication to compete with other scholars who may only focus on publishing nationally or internationally. Thus, the overlap rates between the two groups of Chinese most productive authors are not high in this study because of different publication patterns of Chinese scholars.

For example, 355,387 Chinese authors published 321,875 VIP papers while 123,839 Chinese authors published 36,836 WoS papers in *Pharmacology & Pharmacy*[5] between 2008 and 2015. 103 Chinese authors who published 73 and more VIP papers and 104 Chinese authors who published 42 and more WoS papers were respectively retrieved as Chinese most productive authors from VIP and WoS while 16 scholars were included in both group. 87 out of 103 VIP most productive authors (84.47%) also published WoS papers while 101 out of 104 WoS most productive authors (97.12%) also published VIP papers. Although most of Chinese most productive authors published papers in both national journals and international journals, they have different publication patterns as shown in Figure 8. Some scholars (red nodes in Figure 8) published most of their papers in international (WoS) journals; some ones (blue nodes in Figure 8) preferred to diffuse most of their research results in national Chinese journals; 16 scholars (green nodes in Figure 8) could keep the balance and published their manuscripts in both international and national journals. Thus, it is difficult to evaluate China's research performance based on a single database.

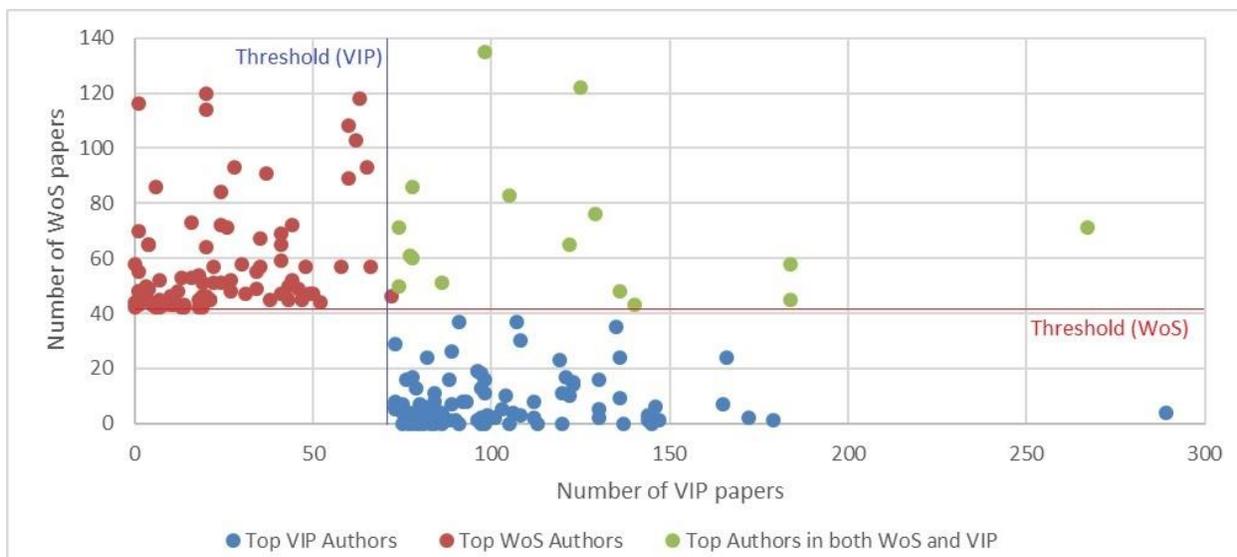

*Figure 8 Publishing Pattern of Chinese Most Productive Authors in Pharmacology & Pharmacy*

---

[5] *Pharmacology & Pharmacy* is selected because its indicators such as China's share in WoS (13.22%), Ratio of WoS papers to all publications (10.27%), Ratio of WoS authors to VIP authors (0.3485), and the overlap rate (15.53%) are close to the average of all disciplines in Natural Science as 14.19%, 10.65%, 0.3678, 13.24% respectively.

### Science Policy

The publication patterns of Chinese scholars are also influenced by China's science policies that promote international publication (Quan, Chen, & Shu, 2017). Since the 1980s, the number of WoS papers has been used to evaluate the research performance in China of both institutions and individuals (Cao, Li, Li, & Liu, 2013; Gong & Qu, 2010) to increase the international visibility of Chinese research. Chinese scholars are required to publish WoS papers to attain promotion, while their affiliated institutions need the number of WoS papers for ranking and funding applications (Cao et al., 2013; YJ Wang & Li, 2015). Chinese research institutions even offer the monetary rewards to their scholars who publish internationally (Quan et al., 2017). These science policies create a negative goal displacement effect (Cao et al., 2013; Frey, Osterloh, & Homberg, 2013; Osterloh & Frey, 2014), the result of which is that, for Chinese scholars, the purpose of publishing their works is not only to advance knowledge but also to fulfill promotion requirements and earn money (Sun & Zhang, 2010; L. Wang, 2016).

In China, international publication is a mandatory requirement for tenure and promotion at most tier-1 universities, but is only an optional requirement at tier-2 universities in which Chinese scholars could use national publications as alternatives (Cao et al., 2013). In order to fulfill the requirement, Chinese scholars from tier-1 universities mostly publish papers in international (WoS) journals while those from tier-2 universities prefer to diffuse their research results in national journals in consideration of the language barrier and high rejection rate of WoS journals. As a result, the proportion of Chinses most productive authors from tier-1 universities is much higher than those from tier-2 universities in WoS, but tier-2 universities contribute the similar number of most productive authors to tier-1 universities in VIP.

### Limitations

A combination of an author's full name and her/his primary affiliated institution is used in this study for name disambiguation. Although this method can disambiguate about 97% of WoS data and almost all VIP data, it cannot disambiguate scholars who were affiliated to different institutions because of academic mobility. In other words, 3% of Chinese scholars are ranked by the number of publications with only one of her/his affiliated institutions, which is a limitation of this study.

In addition, VIP indexes all academic sources in China while WoS only indexes the top international academic journals in each discipline, the difference in terms of the level of coverage between WoS and VIP is also a limitation to this study. We recommend that both a comparison between WoS and CSCD in all Natural Sciences disciplines and a comparison between WoS and CSSCI in all Social Sciences and Humanities should be proposed in future research.

## Conclusion

This study indicates that Chinese scholars do not have homogeneous publication patterns. While some Chinese most productive authors mostly publish in international (WoS) journals, others prefer to diffuse their research results in national Chinese journals. Unsurprisingly, disciplines that are most international in scope such as those of the Natural and Medical

Sciences exhibit a much higher level of overlap than those of the Social Sciences and Humanities. On the whole, these results suggest that the WoS does not accurately represent Chinese research activities based on the extent of coverage of the literature, which confirms the findings of previous research (Guan & He, 2005; Jin & Rousseau, 2004; Jin et al., 2002; Liang et al., 2001; Moed, 2002b; Zhou & Leydesdorff, 2007). However, this study also finds a relative overlap with the Chinese national scientific literature in the Natural Sciences including Life Sciences & Biomedicine, Physical Sciences, and Technology, in which WoS may be used to accurately evaluate Chinese research performance.

1. In Social Sciences and Humanities, in which Chinese scholars publish few WoS papers compared to the large number of publications in national journals, WoS does not represent Chinese research activities. Instead, Chinese bibliometric databases should be used to evaluate Chinese research performance, as suggested by Moed (2002b) and Liang (2003).

2. In Natural Sciences including Life Sciences & Biomedicine, Physical Sciences, and Technology, in which Chinese scholars diffuse their research results in both international journals and national journals, Chinese research performance could be evaluated using a combination of WoS and national Chinese bibliometric databases.

3. In some exceptional disciplines, in which Chinese scholars publish few papers in national journals compared to the large number of WoS papers, national publications cannot represent Chinese research activities when international publications become dominant. In such cases, WoS could be used to evaluate Chinese research performance.

This study also reveals different publication patterns among Chinese scholars: those from tier-1 universities prefer publishing in international journals indexed by WoS, while those from tier-2 universities publish more papers in national journals. Although the difference could be partly attributed to the impact of China's science policies that promote international publication, the detailed relationship between publication patterns and science policies should be investigated in future work.